# Comparison of a Head-Mounted Display and a Curved Screen in a Multi-Talker Audiovisual Listening Task

Gerard Llorach[†,1,2], Maartje M.E. Hendrikse[2], Giso Grimm[2], Volker Hohmann[1,2]

[1]Hörzentrum Oldenburg GmbH, Oldenburg, Germany

[2]Auditory Signal Processing and Cluster of Excellence "Hearing4all", Dept. of Medical Physics and Acoustics, University of Oldenburg

## Abstract

**Introduction:** Virtual audiovisual technology and its methodology has yet to be established for psychoacoustic research. This study examined the effects of different audiovisual conditions on preference when listening to multi-talker conversations. The study's goal is to explore and assess audiovisual technologies in the context of hearing research.

**Methods:** The participants listened to audiovisual conversations between four talkers. Two displays were tested and compared: a curved screen (CS) and a head-mounted display (HMD). Using three visual conditions (audio-only, virtual characters and video recordings), three groups of participants were tested: seventeen young normal-hearing, ten older normal-hearing, and ten older hearing-impaired listeners.

**Results:** Open interviews showed that the CS was preferred over the HMD for older normal-hearing participants and that video recordings were the preferred visual condition. Young and older hearing-impaired participants did not show a preference between the CS and the HMD.

**Conclusions:** CSs and video recordings should be the preferred audiovisual setup of laboratories and clinics, although HMDs and virtual characters can be used for hearing research when necessary and suitable.

## Introduction

In recent years, audiovisual technologies have become more prevalent in hearing research [1]–[11]. One of the motivations of using such technologies is to increase the ecological validity of the experiments in the laboratory and the clinic [12], i.e., that the results in the laboratory are representative of situations in the real world [13], [14]. This is particularly important when fitting hearing aids for the first time. New users tend to give up using hearing aids if these don't improve their hearing situation in their daily environments, thus leading to a poorer quality of life in the long term [15], [16].

Audiovisual technologies such as head-mounted displays (HMDs) and surrounding screens are already established and are available in the market. For hearing and hearing aid research, however, the applicability and acceptance of different audiovisual technologies has hardly been investigated. [1] tested speech perception with and without HMDs and asked the participants about technology preference and its applicability in the clinic. Almost all participants were willing to complete the test in a clinical setup with the HMD, but the weight of the device and the participant's unfamiliarity with it were concerning issues. Seol et al. [1] mentioned it is crucial to test and validate the audiovisual technology used in audiological experiments, "as it could be one of many factors that professionals and patients would consider before employing and performing the test in clinics". More data are therefore needed to characterize and establish audiovisual technology options in hearing research [17].

Individual characteristics, in particular age and hearing type, could elicit different technology acceptance. Most research in technology acceptance has been done with young normal hearing (YNH) participants, thus data is lacking for older and hearing impaired participants. Philpot et al. [18] found no difference in preference for young adults between the CS system and the HMD when watching a 360-degree angle documentary. Hendrikse et al. [19] found that YNH participants preferred video recordings over animated virtual characters in a listening task with a curved screen. Older and hearing-impaired participants might be more reluctant than YNH participants to use intrusive and immersive audiovisual technologies, although the opposite is possible as well. Surrounding screens and HMDs differ in several characteristics that could affect acceptance in laboratory measurements. A HMD is worn on the head and it occludes all visual references to the real space. The field of view is reduced, as current consumer-grade HMDs cannot cover the whole human field of view (210º horizontal). The HMD requires head straps that could interfere with the positioning and performance of the hearing aids, and therefore could be uncomfortable for hearing aid users. Additionally in [1], HMDs were reported as heavy and difficult to use if one is not familiar with the device. With surrounding screens, the real space is always visible. The user's head movement and vision are less constrained, as no device is worn on the head. To understand such differences between displays and visual conditions, the current study provides comparative data for different readily available audiovisual technology options, in particular for curved screens (CSs) vs. head-mounted displays (HMDs), and for video recordings (VID) vs. virtual characters (VC) vs. audio-only (AO). Technology acceptance is particularly important for the applicability of such technologies in clinical setups. If we want to use such technologies in the clinic, we must make sure that older and hearing-impaired participants are willing to use such technologies and that these technologies do not deter the participants from getting their hearing abilities checked. In consideration of that, this study included YNH participants, older normal hearing participants (ONH), and older hearing-impaired participants (OHI).

In this experiment, participants listened to conversations in different audiovisual conditions and answered questions about the content afterwards; subjective ratings and open comments of preference and technology acceptance were collected and analyzed. The results of this study are meant to provide useful insight to guide future research and implementation in hearing clinics and research laboratories using audiovisual technologies.

This study replicates parts of the experimental setup of our previous study Hendrikse et al. (2018b). The current work extends it by adding the HMD as a display type, by testing older participants, with and without hearing impairment, and by measuring technology acceptance with open interviews. Head orientation and gaze were measured in the experiment, but the analysis and results are to be presented in a future article. A direct comparison between several audiovisual setups is presented here, as two display types were combined with three visual conditions. Additionally, older and older hearing-impaired participants were included to compare the applicability of the audiovisual setups in clinical environments.

## METHODS

The participants were asked to listen to conversations under six different conditions: two display types combined with three visual conditions. The display types were a CS and a HMD, and the visual conditions were audio-only, virtual characters, and video recordings. In Fig. 1, the conditions can be seen as they looked in the experiment. The top row shows the conditions with the HMD and the lower row the ones with the CS. The task of the participants was to answer three questions about the content of the conversation they just heard. After completing all six conditions, they had to do an interview and fill out questionnaires. The CS-AO condition was meant to represent an experiment in a hearing laboratory without visual cues, which is the standard case in hearing research. To counterbalance the number of conditions done with the CS, the HMD-AO condition was included.

## Participants

Seventeen young normal-hearing subjects (YNH), ten older normal-hearing subjects (ONH), and ten older moderately hearing-impaired subjects with hearing aids (OHI) participated in the study. All but one of the YNH subjects were students of the Carl von Ossietzky Universität Oldenburg with a mean age of 24 years (STD 2.43, range 18-27). YNH subjects were specifically asked about their hearing: none of them reported hearing loss. The mean age of the older participants was 61.9 years (STD 5.3, range 50-69). ONH and OHI participants were recruited through Hörzentrum Oldenburg GmbH, where their audiograms were measured regularly: ONH participants had a mean pure tone average (PTA) between 125 Hz and 8 kHz of 10 dB HL; the mean PTA between 125 Hz and 8 kHz for OHI participants was 49.4 dB HL. The OHI participants had been using their hearing aids for more than six months and had a moderate symmetric hearing loss. They wore their hearing aids during the experiment. Participants were also specifically asked about visual impairments, which none of them reported (e.g., reduced vision not corrected by glasses or contact lenses). The ethics permission was granted by the ethics committee of the CvO Universität Oldenburg (Drs. 1r63/2016). The participants signed an informed consent.

## Setup

The experiment was conducted inside a circular 'tent' within an acoustically semi-treated room (reverberation time (T60) = 0.13s). The inside of the tent and a top view of the room can be seen in Fig. 2. The figure shows where the participant was sitting and how the projection looked for the CS-VID condition. The position of the elements of the tent is also shown in Fig. 2. The tent was covered with a black blanket and it had a radius of 1.98 meters. It consisted of a metal structure that supported a circular array of 16 loudspeakers (Genelec 8020B, Genelec Oy, Olvitie, Finland) and an acoustically transparent curved screen. The loudspeakers were spaced every 22.5-degree angle at a radius of 1.96 meters and a height of 1.60 meters. The curved screen was in front of this array of loudspeakers and was 2 meters tall with a 1.76-meter radius. Images were projected onto the screen from a close-field projector (NEC U321H, Sharp NEC Display Solutions, Munich,

Germany) placed on top of the tent. The projector achieved a projection of 120-degree angle (horizontal) and had a refresh rate of 60 Hz with a resolution of 1920x1080 pixels. The HTC Vive system (HTC Corporation, New Taipei City, Taiwan) was used as HMD. The HTC Vive Base Stations and a camera for live feedback were placed above the curved screen. The HTC Vive display had a refresh rate of 90 Hz, a resolution of 1080x1200 pixels per eye, a 100-degree angle field of view (horizontal) and orientation and translation tracking. The background noise level inside the tent with all the devices working was 31.1 dB A.

A chair was placed in the center of the tent, facing towards the front, i.e., the 0-degree angle azimuth of the simulation. The chair was on an elevated platform with dimensions 120 cm by 120 cm. The platform was elevated 30 cm from the floor. When the participants were seated, the ears were at approximately the same level as the loudspeakers (1.60 meters). To the side of the participant, around 120-degree angle azimuth from the front, there was an emergency button at arm's reach: pressing this button stopped the simulation.

Three computers were used in the experiment: an Ubuntu 14.04 for the acoustic rendering, data logging and master control; an Ubuntu 14.04 for the screen projection with NVIDIA Quadro K6000; and a Windows 10 for the HMD rendering with NVIDIA Quadro M5000 and head tracking.

The 3D virtual acoustic environment was rendered with TASCAR (Grimm et al. 2019b) versions 0.175.2-0.177.5. The virtual 3D scene for the curved screen was created and rendered with the Blender Game Engine version 2.79 (Roosendaal 1995). The image warping for the projection was done with the graphics card and was manually configured and calibrated. The 3D scene for the HMD was rendered with the Unity game engine version 2017.1.0f3. All the sensor data was transmitted for central data logging in TASCAR via the Open Sound Control (OSC) (Wright and Freed 1997) and the LabStreamingLayer protocol (Kothe et al. 2018). The experiment was controlled and executed with Matlab 2016b and with the acoustic engine using the OSC protocol. Temporal alignment between visual and acoustic cues was adjusted manually.

Head orientation was measured with two different devices for the CS and the HMD. For the CS, participants wore a head crown with a Vive Tracker (HTC Corporation, New Taipei City, Taiwan) attached. For the HMD, the device itself, i.e., the HTC Vive, was used for head tracking. The horizontal movement of the eyes was measured with two electrodes placed next to the eyes (electrooculography, EOG).

## Stimuli

We used the same audiovisual material, casual acted conversations, as in our previous study (Hendrikse et al. 2018b). There are seven conversations available, one of which we used for the training and the six remaining for the experiment conditions. The material can be found in the database by Hendrikse et al. (2018a). The conversations lasted between 1 min 24 s and 1 min 39 s and the topics were food, holidays/travelling, weather, work, future plans, movies and anecdotes. Of the four talkers, two were females and fluent non-native talkers (German CEFR C1), and the other two were males and native talkers. In the 3D virtual scene, the actors were positioned at 45, 15, -15 and -45-degree angle in a radius of 1.7 meters away from the listener's position. After each conversation, one of the actors asked three multiple-choice questions about the content. In this experiment, we did not record the answers of the questions but this was unknown to the participants in order to keep them engaged. These questions can be found in the aforementioned database (Hendrikse et al. 2018a).

### *Acoustic Stimuli*

The acoustic conditions were the same across all trials. The multi-talker conversations were played together with diffuse background noise. In our laboratory, the loudspeaker layout did not match the position of the target talkers (see Fig. 2). We used TASCAR to generate a virtual acoustic environment and to reproduce sound sources at the prescribed place. This virtual acoustic environment simulated a virtual source for each target talker at the predefined position, as if the clean speech was played back through a loudspeaker in the room at the respective position. The audio reproduction technique used for the target talkers was horizontal 7th-order Ambisonics with max-rE decoding (Daniel et al. 1998), rendered to the 16 loudspeakers on a ring at ear level. The diffuse background noise was a 1st-order Ambisonics recording of the cafeteria of the University of Oldenburg (Hendrikse et al. 2018a). To achieve a diffuse reproduction of the background sound field and to avoid spectral artifacts due to self-motion, the first-order signal was extended to $7^{th}$ order, and a frequency-dependent rotation similar to the method of Zotter et al. (2014) was applied. The average sound levels for each conversation were measured with a sound level meter at the position of the listener. The sound levels for the YNH were 45.2 ± 0.3 dB A for the conversations and 49.7 dB A for the cafeteria background noise. For the older participants, the speech levels had to be increased and the noise levels reduced, as the first two older participants complained that they could not hear the spoken instructions clearly inside the simulation (speech in quiet). The levels for the ONH were adjusted with an increase of 3.1 dB for speech (48.3 dB A) and a decrease of 3.7 dB for noise (46.0 dB A). The levels for the OHI were adjusted with an increase of 9 dB for speech (54.2 dB A) and a decrease of 6 dB for noise (43.7 dB A). These level adjustments were defined by the first participants (first ONH and first OHI). The speech level was raised so that they could understand the speech in quiet and the background noise level was reduced so that the conversation could be followed when background noise was present. We were aiming for a realistic speech level around 65 dB A, but due to a calibration error, the actual speech levels were not in this range.

### *Visual Stimuli*

Three different visual conditions were presented in this experiment (see Fig. 1): audio-only (AO), virtual characters (VC) and video recordings (VID). In the CS-AO condition, the projection was turned off and a diffuse light was turned on. In the HMD-AO condition, a virtual laboratory was shown, so the participant would feel he/she was in the same real space and would have some reference points: the participant could see the chair underneath, the platform where the chair was, the cylindrical screen and the emergency button. This virtual scene was used for the other visual conditions. For the VC condition, the 3D virtual characters were created with Makehuman version 1.02 in resemblance to the real actors. The virtual characters were blinking and moving their lips with a speech-based lip-syncing (Llorach et al. 2016). The virtual characters also moved their head and eyes: they followed the conversation by looking towards virtual character who was speaking. These three animations were automated and generated in real-time. The effects of these animations can be found in the studies by Grimm et al. (2019b) and Hendrikse et al. (2018b). In the VID condition, the video recordings were shown through flat screens in the virtual scene (see Fig 1).

### *Experiment Procedure*

The participants filled in an anonymization form and an informed consent. They were informed about the experiment through written forms, a video clip and orally. The interpupillary distance was measured with a ruler and the lenses of the HMD were adjusted accordingly. The head crown and the HMD were adjusted to the participant's comfort. If the participants used corrective glasses, we let them try the HMD with and without them; they decided whether they wanted to do the HMD trials with or without glasses. The EOG electrodes were attached to the participant together with a Bluetooth transmitter and participants were instructed not to touch them during the experiment. They were instructed that they would have to answer verbally 'A', 'B' or 'C', to the multiple-choice questions presented after each conversation. After this introduction, they filled out the pre-exposure Simulator Sickness Questionnaire (SSQ; Kennedy et al. 1993) and were seated on the chair inside the tent. We included the SSQ in the experiment to assess whether participants suffered from cybersickness.

The participants started with the HMD or the curved screen randomly. They did the three randomized visual conditions with one display followed by three more with the other display. The order of the visual conditions was the same with the curved screen and the HMD for each participant. The conversations were randomized and each conversation was played equally often for each condition across all participants. The EOG required a calibration protocol, which was done once for the CS and once for the HMD before starting the trials.

Instructions about the task were repeated through a virtual character in the simulation. When using the HMD, an initial adaptation phase was added: a virtual character made suggestions for getting used to the room, to look at the chair they were sitting on and to find the emergency button behind them. If they did not find the emergency button, the researcher came inside the tent and made sure the participant could turn and see the button. The virtual button was in the same location as the physical one. This procedure was done to adapt the participants to the experience, e.g., some participants may be unaware that they can move or turn their heads with the HMD. This adaptation phase lasted around 1 minute.

After the instructions, there was a training trial. The training trial used a conversation that was not used in the test trials. After each conversation, the participants answered verbally to the multiple-choice related questions. The participants came out of the tent to fill out the SSQ after all trials were completed. After this, we proceeded with the open interview recorded with a sound recorder.

### Measures

The preference and acceptance of the audiovisual conditions were measured via a recorded interview. The participants were asked to give comments and impressions about the experiment once they completed all listening tasks. They were given a paper with six pictures (one for each condition) and a picture of each display device. We allowed a minimum of three minutes time and a maximum of 15 minutes for comments. Afterwards, the participants were asked to select one of the six conditions (see Figure 1) as the one they would like to experience in a future experiment. Then, they were asked to name the second-best condition. Finally, they were asked to choose if there was any condition they would not like to experience again. The participants that did not have a preference between displays or visual conditions could also answer combinations, i.e., first preference as the video regardless of the display. The increase in SSQ symptoms between pre- and post-exposure questionnaire was computed and the mean values for the total simulator sickness severity were below 13 for both groups. According to Kennedy et al. (1993), the cybersickness reported in this experiment is considered insignificant (10-15 Total Severity).

## Results

### Open Comments

We analyzed the recorded interviews and annotated the issues that were mentioned: these are summarized in Table 1. The interviews revealed that the speech was difficult to understand (Table 1.2); some subjects found the males talkers more difficult to understand (Table 1.3-5); some found the accent of the non-native female talkers hard to understand (Table 1.6). Three participants mentioned that moving their head changed their audio perception (Table 1.7). Five participants mentioned that the HMD was heavy and three older participants commented that they felt isolated when wearing the HMD (Table 1.8-9). Three YNH participants noticed that the screen of the HMD was brighter than the CS (Table 1.10). Seven participants mentioned that in the AO trials it was easier to concentrate than in the other trials, but for three participants it was the opposite (Table 1.12-13). Additionally, eight participants mentioned that it was easier to understand the conversation in the VID condition (Table 1.14). Four OHI and two ONH participants complained about the insufficient resolution of the lips of the virtual characters (Table 1.16), four participants mentioned that the virtual characters were too stiff (Table 1.17) and seven participants indicated that the characters were not realistic (Table 1.15).

### Chosen conditions

Participants were asked to select, out of the six possible conditions, the two most preferred, and to mention whether there was any they would not like to repeat. The participants could also choose a visual condition, e.g., VID, regardless of the display or vice versa. The first four subjects were not asked whether there was any condition they would not like to do again. The answers of the participants are shown in Table 2. We divided the preference results by visual conditions and display. The first and second preferences were grouped together, e.g., the VID condition was chosen by thirty-four participants out of thirty-eight as the first and/or second preference. Eighteen participants out of thirty-four were willing to do all the conditions again.

The VID condition was clearly chosen as the preferred visual condition and was never rejected. The other two visual conditions, VC and AO, were chosen with nearly equal preference. The YNH and the OHI participants showed no preference between the HMD and the CS displays. The ONH preferred the CS more often (all ONH participants chose the CS and six chose the HMD as first/second condition out of ten). In general, the HMD was more frequently rejected by the YNH and the OHI and the CS was rejected only by two participants out of thirty-three. The rejected conditions were always a combination of a display (HMD or CS) with the AO or VC condition. The AO condition was rejected by four participants more than the VC condition (eleven vs. seven participants out of thirty-three). No ONH participants rejected the AO condition, whereas almost half of the YNH and the OHI rejected it.

## Discussion

As expected based on the study by Philpot et al. (2017), the YNH participants showed equal preference for the two displays. The ONH participants preferred the CS over the HMD, but almost equal preference was shown for the OHI participants. The HMD was rejected more often as a display and received more negative comments, such as that it was heavy, isolating and distracting. Therefore, the CS would be a better choice for the comfort of the participants. Nevertheless, the HMD was chosen quite often as a first or second option and it was never rejected as a display alone, i.e., regardless of the visual condition. Most OHI were willing to use the device for future experiments. Therefore, it should be considered for clinical implementations, as a cheaper and simpler implementation.

The video recording condition (VID) was clearly the most preferred visual condition. This finding agreed with the previous study of Hendrikse et al. (2018b). The AO condition was the most rejected (by eleven participants out of thirty-three), showing a general preference for conditions with visual cues. The comments regarding the VC condition indicated that their quality, non-verbal behaviors and lip-readability should be improved. It is worth noticing that only the older participants (two ONH and four OHI participants) mentioned the lip-readability, indicating that older participants might look for this kind of visual cues specifically.

The accent was brought up as a difficulty for understanding, but the female talkers, which were the ones with the accent, were also specified to be easier to understand by other participants. We consider that the talkers were all equally intelligible, according to the open comments.

### Outlook and Limitations

Depending on the research question, current hearing clinics and laboratories might want to use immersive visual cues (Keidser et al. 2020). Changing their methodologies from audio-only to audiovisual stimuli might be expensive and effortful. HMDs are more affordable and easier to setup than custom-built CSs. Nevertheless, specific procedures need to be done in the clinic for HMDs, such as measuring the interpupillary distance and adjusting the head-straps. Additionally, head-mounted displays introduce acoustic distortions [20], [21] that might affect the results collected in the clinic. Which audiovisual system to use will depend on each specific experiment and clinical setup.

Next generations of HMDs might improve some of the issues mentioned by the participants, such as the weight of the device. Mixed reality and augmented reality solutions should also be considered, as they might be less isolating than HMDs.

Further improvements need to be done to our virtual characters if they are to be used (see Llorach et al. 2018). This is supported by the comments of the participants and the significant differences found between the video recordings and the virtual characters. In this study we used open-source characters and animations available at the time.

## Acknowledgments

Thanks to: K. Schwarte for her help during the measurements and the analysis of the recorded conversations; M. Krüger and M. Zokoll for the support on recruiting older participants; A. Wagner for counseling; and J. Luberadzka for the helpful comments. This study was funded by the European Union's Horizon 2020 research and innovation programme under the Marie Sklodowska-Curie grant agreement No 675324 (ENRICH) and the Deutsche Forschungsgemeinschaft (DFG, German Research Foundation) - Projektnummer 352015383 - SFB 1330 B1.

## Data Availability Statement

The audiovisual stimuli associated with this article is available in Zenodo ("Audiovisual recordings of acted casual conversations between four speakers in German"), under the reference https://doi.org/10.5281/zenodo.1257333.

## Figures and Legends

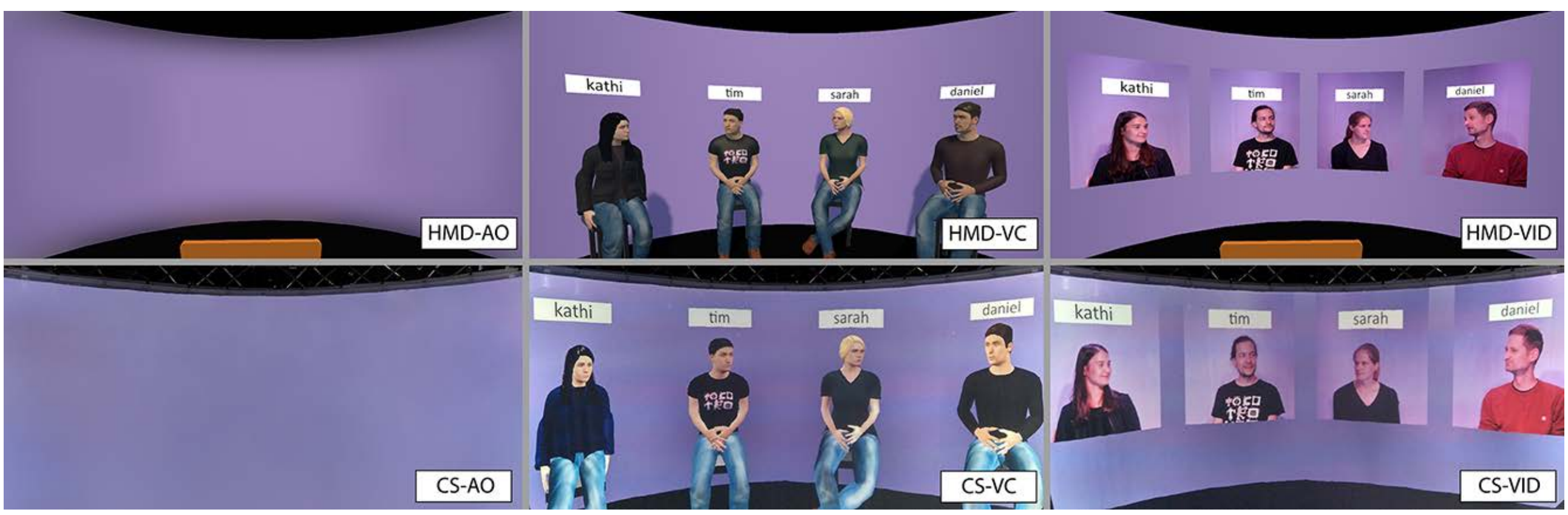

**Figure 1: Images of the six conditions presented in this experiment. The images on the top row are screen captures of the virtual space used for the head-mounted display (HMD). The virtual space contained elements of the real space such as the chair where the participants were seated. The images on the bottom row are pictures of the curved screen (CS) taken inside the laboratory. From top to bottom: conditions with the HMD and the curved screen. From left to right: audio-only (AO), virtual characters (CS), and video recordings (VID).**

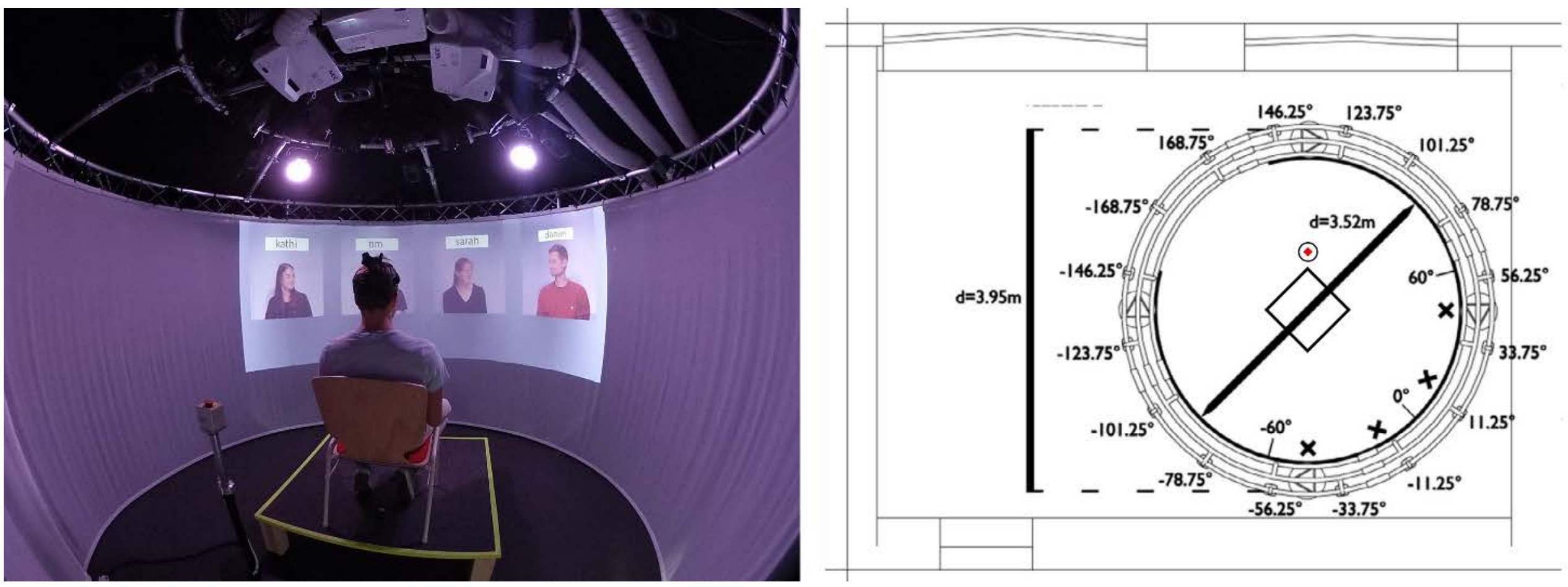

**Figure 2: A, On the left: fish-eye picture of the inside of the tent in the condition with the curved screen and the video recordings. B, On the right: top view of the tent and the room. The angles on the outside of the metal ring (circular structure) indicate the position of the loudspeakers. The crosses indicate the position of the target speakers in this experiment. The square in the middle represents the platform where the participant was seated. The circle with a red dot, close to the platform, depicts the emergency button.**

## Tables

**Table 1. Comments by the participants during the open interviews. Only comments mentioned by three or more participants were noted in this table.**

| | Nº of YNH out of 17 | Nº of ONH out of 10 | Nº of OHI out of 10 | Total nº of participants out of 37 |
|---|---|---|---|---|
| **Comments about the conversations and the acoustics** | | | | |
| 1. It was hard to concentrate | 2 | 1 | 3 | 6 |
| 2. It was difficult to understand | 7 | 4 | 4 | 15 |
| 3. It was easier to listen to the female talkers | 3 | 2 | 0 | 5 |
| 4. Daniel (+45-degree angle) was really hard to understand | 0 | 2 | 2 | 4 |
| 5. Tim (-15-degree angle) was really hard to understand | 2 | 2 | 0 | 4 |
| 6. It was hard to understand the accent | 1 | 2 | 1 | 4 |
| 7. The head position changed the audio perception | 1 | 2 | 0 | 3 |
| **Comments about the display** | | | | |
| 8. I felt isolated with the head-mounted display (HMD) | 0 | 1 | 2 | 3 |
| 9. The HMD was heavy | 2 | 1 | 2 | 5 |
| 10. The image was brighter with the HMD | 3 | 0 | 0 | 3 |
| 11. Wearing the HMD was distracting | 1 | 2 | 0 | 3 |
| **Comments about the visual condition** | | | | |
| 12. It was easier to concentrate in the audio-only (AO) condition | 4 | 1 | 2 | 7 |
| 13. It was harder to concentrate in the AO condition | 3 | 0 | 0 | 3 |
| 14. It was easier to listen to the video recordings | 5 | 3 | 0 | 8 |
| 15. The virtual characters (VCs) were not realistic | 4 | 0 | 3 | 7 |
| 16. The lips were not readable with the VCs | 0 | 2 | 4 | 6 |
| 17. The VCs were too stiff | 3 | 0 | 1 | 4 |

**Table 2. Preferences for the visual conditions and displays.**

| | | **Chosen as 1st or 2nd condition** | | | |
|---|---|---|---|---|---|
| | | Nº of YNH out of 17 | Nº of ONH out of 10 | Nº of OHI out of 10 | Total nº of participants out of 37 |
| **Visual condition** | Video recordings | 15 | 9 | 9 | 33 |
| | Virtual characters | 6 | 4 | 3 | 13 |
| | Audio-only | 5 | 3 | 3 | 11 |
| **Display** | Head-mounted display | 15 | 6 | 8 | 29 |
| | Curved screen | 15 | 10 | 9 | 34 |
| | | **Never again condition** | | | |
| | | Nº of YNH out of 13 | Nº of ONH out of 10 | Nº of OHI out of 10 | Total nº of participants out of 33 |
| **Visual condition** | Video recordings | 0 | 0 | 0 | 0 |
| | Virtual characters | 3 | 1 | 3 | 7 |
| | Audio-only | 6 | 0 | 5 | 11 |
| **Display** | Head-mounted display | 5 | 1 | 4 | 9 |
| | Curved screen | 1 | 0 | 1 | 2 |